\documentclass[10pt,a4paper]{article}
\usepackage{amsmath,amssymb,amsfonts,amsthm}
\usepackage{graphicx}

\newtheorem{theorem}{Theorem}[section]
\newtheorem{lemma}[theorem]{Lemma}

\newcommand{\e}{\varepsilon}

\newcommand{\Ldt}{\hat\Delta}

\newcommand{\Ld}{\bar\Delta }

\newcommand{\Rt}{\mathbb{R}^3}

\newcommand{\orden}[1]{\mathcal{O}\left(#1\right)}

\newcommand{\rt}{\tilde r}

\title{The Yamabe invariant for axially symmetric two Kerr black
  holes initial data}

\author{Gast\'on A. Avila$^1$ and Sergio  Dain $^{1,2}$\\
\\
$^1$Facultad de Matem\'atica, Astronom\'{i}a y F\'{i}sica, \\
Universidad Nacional de C\'ordoba,\\
 Ciudad Universitaria, (5000) C\'ordoba, Argentina.  \\
\\
$^{2}$Max Planck Institute for Gravitational Physics,\\
  (Albert Einstein Institute), Am M\"uhlenberg 1,\\
  D-14476 Potsdam Germany. }

\begin{document}

\maketitle

\begin{abstract}
  An explicit 3-dimensional Riemannian metric is constructed which can
  be interpreted as the (conformal) sum of two Kerr black holes with
  aligned angular momentum. When the separation distance between them
  is large we prove that this metric has positive Ricci scalar and
  hence positive Yamabe invariant. This metric can be used to
  construct axially symmetric initial data for two Kerr black holes
  with large angular momentum.
\end{abstract}

\section{Introduction}

%Initial data for binary black holes: physical relevance

The numerical study of the binary black hole problem has recently
made significant progress \cite{Pretorius:2005gq, Campanelli:2005dd,
  Baker:2005vv}. It is now possible to calculate the evolution of a wide
variety of astrophysical scenarios. In order to do this, appropriate 
initial data for the Einstein equations are necessary, such that they 
describe the initial conditions of these scenarios.

% The conformal method and the yamabe number: limitation of the spin

A natural physical requirement for binary black hole initial data is
that in the limit of large separation distance each of the black holes
approximates an stationary, isolated, black hole (in vacuum, this
implies that the black hole approximates the Kerr black
hole). Otherwise, the data will contain spurious radiation which can
in principle contaminate the final waveform (see
\cite{Hannam:2006zt}).

The nonlinearity of the constraint equations makes the problem of
constructing initial data satisfying such far limit a nontrivial one.
In \cite{Dain00c} such data for two Kerr black holes with far limit
has been constructed. The procedure uses the conformal method for
solving the constraint equations (see the review article
\cite{Bartnik04b} and references therein). In this construction, one
starts by constructing a superposition of two conformal Kerr metrics
to get a new conformal metric, which is then used as a `seed' metric
for the conformal method. Then, a conformal factor is calculated such
that the corresponding rescaled metric satisfies the constraint
equations. The existence and uniqueness of that conformal factor can
be proved provided the `seed' metric satisfies an important
requirement: its Yamabe invariant must be positive (see
\cite{Bartnik04b} for a discussion about the Yamabe invariant in the
context of the constraint equations). In order to ensure this
condition in \cite{Dain00c}, it has been assumed that the individual
angular momentum parameters of the black holes are small with respect
to the masses.

% Physical relevance of large spins
This assumption is, of course, an undesirable restriction. Highly 
spinning black holes are relevant in many astrophysical situations. A 
remarkable example of this are the large merger recoil kicks (see 
\cite{Herrmann:2007ac,Campanelli:2007ew, Koppitz:2007ev, Dain:2008ck}).

% The results obtained here: far limit. 
The purpose of this article is to overcome this restriction for the
axially symmetric case. We provide an explicit metric, having three
`ends', which is constructed as a superposition of two Kerr black
holes metrics (with aligned or anti aligned angular momentum). When
the mass parameters of one black hole is zero then the metric reduces
exactly to the Kerr metric. In this sense, we say that the metric
satisfies the far limit. The important property of this metric is that
when the separation distance between the black holes is large enough
its Ricci scalar  is positive.  The parameters of the
individual Kerr black holes are only restricted by the Kerr inequality
$|a|\leq m$, where $a$ is the angular momentum per mass unit and $m$
is the mass of the individual black hole.  
That is, the angular momentum is not assumed to be small. Instead, we
assumed the physical reasonable condition that the separation distance
between them is large. This constitutes the main result of the article.

When the three `ends' are asymptotically flat (the non-extreme
cases) the positivity of the Ricci scalar implies the positivity of
the corresponding Yamabe invariant.  And hence the metric can be used
as a seed metric for the construction mentioned above. In this way
axially symmetric initial data can be constructed for two Kerr-like
black holes with large angular momentum. However, we emphasize that
our result covers also the extreme limit $|a|= m$ of any of the black
holes. We believe that this can be used to construct (the still
unknown) initial data for two extreme Kerr black holes.

\section{Main result}
Consider the Kerr black hole (i.e. the Kerr metric such that $|a|\leq
m$) in Boyer-Lindquist coordinates. Take any constant $t$ slice in these
coordinates and denote by $\tilde h_{ij}$ the intrinsic 3-dimensional
metric of the slice. There exist spatial coordinates $(\rho,z, \phi)$
on the slice such that the metric $\tilde h_{ij}$ has the following
form
\begin{equation}
\label{eq:kerr-metric}
 \tilde h_{ij}=(1+u)^4 h_{ij},
\end{equation}
where 
\begin{equation}
  \label{brilldef}
 h_{ij}=e^{2q} (d\rho^2+dz^2)+ \rho^2 d\phi^2, 
\end{equation}
and the functions $q$ and $u$ (which do not depend on the coordinate
$\phi$) are explicitly given by equations (\ref{eq:51}) in the
appendix \ref{sec:kerr-data}.  The slice is a 3-dimensional manifold
$M$ which has the topology  $M=\Rt\setminus \{0\}$, where $\{0\}$
denotes the origin in the coordinates  $(\rho,z, \phi)$. For the
results presented here a relevant property of the
 Kerr metric is that $u\geq 0$ (see
Lemma \ref{t:psi}). 

We construct a new manifold $S$ by removing two points from $\Rt$
denoted by $s_1$ and $s_2$, namely $S=\Rt\setminus \{s_1,s_2 \}$.  Let
$(\rho,z, \phi)$ be cylindrical coordinates on $S$ such that the two
points $s_1$ and $s_2$ are located at the axis $\rho=0$, separated by
coordinate distance $d$. This requirement does not completely fix the
coordinate system, since there still is a translation freedom in the $z$
coordinate $z\to z + c$ where $c$ is any arbitrary constant.
Later on we will make use off this freedom to simplify the
computations. However, for the formulation of the results there is no
loss off generality if we chose the coordinate system such that $s_1$
is located at $d/2$ and $s_2$ at $-d/2$ on the $z$ coordinate axis. 

Given the functions $q(\rho,z; m_1, a_1)$ and $u(\rho,z; m_1, a_1)$ 
of the Kerr initial data, with parameters $(a_1,m_1)$, we define the new
functions $(q_1,u_1)$ by
\begin{equation}
  \label{eq:4}
  q_1=q(\rho, z-d/2; m_1,a_1), \quad u_1=u(\rho, z-d/2; m_1,a_1).  
\end{equation}
In the same way, for functions $q(\rho,z, m_2, a_2)$ and $u(\rho,z,
m_2, a_2)$ we define $(q_2,u_2)$ by
\begin{equation}
  \label{eq:4b}
  q_2=q(\rho , z+d/2;m_2, a_2  ), \quad u_2=u(\rho, z+d/2;m_2, a_2).  
\end{equation}

With this set of functions constructed from the Kerr metric, we define
a metric on $S$, depending on the parameters
$(d,m_1,a_1,m_2,a_2)$, by
\begin{equation}
 \label{brillsuma}
 \tilde H_{ij}=(1+u_1+u_2)^4 H_{ij}, 
\end{equation}
where
\begin{equation}
  \label{eq:2}
 H_{ij}= e^{2(q_1+q_2)}(d\rho^2+dz^2)+\rho^2 d\phi^2.
\end{equation}
Since $u_1$ and $u_2$ are positive functions, $\tilde H_{ij}$ is well
defined as a metric on $S$ for any choice of the parameter $d$. 

For the non-extreme case (i.e $|a_1| < m_2$, $|a_2| < m_2$) one can
check that the metric $\tilde H_{ij}$ is asymptotically flat with
three asymptotic ends. These ends are denoted by $s_\infty$, which
correspond to the limit $r\to \infty$ ( $r$ denotes the euclidean
distance to the origin) and $s_1$, $s_2$, which
correspond to the limits $r_1\to 0$ and $r_2\to 0$ respectively ($r_1$
and $r_2$ denote the euclidean distances with respect to the points
$s_1$ and $s_2$ respectively).  The
total mass of the metric (\ref{brillsuma}) at the end $s_\infty$ is
given by $m_1+m_2$.

If one of the ends is extreme (that is $|a_1|=m_1$ or  $|a_2|=m_2$),
then the metric will still be asymptotically flat at the end
$s_\infty$ but will not be asymptotically flat at the extreme end. 

Denote by $R_{\tilde H}$ the Ricci scalar of the metric
(\ref{brillsuma}). 
The following theorem constitutes the main result of this article. 
\begin{theorem}
\label{t:main} 
There exists a constant $d_c$ such that if $d\geq d_c$ 
then $R_{\tilde H}\geq 0$ on $S$. Moreover, if $a_1$ or $a_2$ is
different from zero, then $R_{\tilde H}> 0$ on $S$.
\end{theorem}
For $a_1=a_2=0$ we have $R_{\tilde H}=0$. In this limit the metric
reduces to the Brill-Lindquist metric \cite{Brill63}. If we set to
zero the mass of one of the black holes (this implies that the
corresponding angular momentum is also zero) then the metric
(\ref{brillsuma}) reduces to the Kerr metric (\ref{eq:kerr-metric}).

Consider the Yamabe invariant defined on the manifold $S$ for the
metric $\tilde H_{ij}$ 
\begin{equation} 
 \label{eq:yamabe} 
\lambda= \inf_{\varphi\in C^\infty_c(S), \, \varphi \not \equiv0}
\frac{\int_S (8|D\varphi|^2 +  R_{\tilde H}\varphi^2)d\mu }{ ( \int_S
  |\varphi|^6 d\mu)^{1/3}} ,
\end{equation}
where $C^\infty_c(S)$ denotes the set of smooth functions with compact
support in $S$, $|D\varphi|^2=\tilde
H^{ij}\partial_i\varphi\partial_j\varphi$, $d\mu$ is the volume
element with respect to $\tilde H_{ij}$, and $\varphi \not\equiv 0$
means that $\varphi$ cannot be identically zero everywhere.

For the non-extreme cases, theorem \ref{t:main} implies that $\lambda
> 0$ (it is obvious that $\lambda \geq 0$, to prove that it is
strictly positive we use Lemma 4.1 in \cite{Cantor81b}). 

The main application of theorem \ref{t:main} is the construction of a
two (non-extreme) Kerr black holes data. The results proved so far
assumed that the angular momentum parameter $a$ is small with respect
to the mass (see \cite{Dain00c}). Theorem \ref{t:main} allows to
generalize these results for any $a$ such that $|a|< m$, but only for
the axially symmetric case (that is, when the angular momentum of the
black holes are aligned or anti aligned).  Using the conformal method
and the metric (\ref{brillsuma}) as a `seed' conformal metric the
complete data can be obtained as follows.  Since this procedure
involves standard applications of known results we only sketch the
proof.  First we superpose two extrinsic curvatures from two Kerr
initial data. This can be done in general (see \cite{Dain00c}),
however in axial symmetry the procedure is much simpler and can be
done explicitly (see, for example, \cite{Dain99b},
\cite{Dain06c}). Then, we use existence theorems for the
Lichnerowicz equation (see for example, \cite{Maxwell04},
\cite{Choquet99}, \cite{Dain00c}) to prove that there exists a new
conformal factor such that a conformal rescaling of the metric $\tilde
H_{ij}$ satisfies the vacuum constraint equations with the above
constructed second fundamental form (see \cite{Dain00c} for details).
It is in this last step when theorem \ref{t:main} plays an important
role. In order to apply these existence theorems we need to ensure
that the conformal metric has positive Yamabe invariant. 
Using theorem \ref{t:main} we
only require  that the separation
distance between the black holes is large.  We also note that although
this procedure is identical to the one proposed in \cite{Dain00c}, the
`seed' conformal metric is different.

The above construction applies to the non-extreme case. Remarkably,
theorem \ref{t:main} is valid also for the extreme case $|a|=m$. It is
very likely that this theorem can also be used to construct a
superposition of two extreme Kerr black holes. But this remains to be
seen. The existence results for the Lichnerowicz equations proved so
far in the literature applies only to asymptotically flat manifolds and
not to manifolds with cylindrical ends like extreme Kerr.

\section{Properties of Kerr initial data}
\label{sec:prop-kerr-init}
In this section we establish the three key properties of Kerr
intrinsic metric which allow us to prove theorem \ref{t:main} in the
next section. These properties are collected in lemmas \ref{t:psi}, 
\ref{lemmaRtildePositivo} and \ref{lemmaRmenosCompacto}.

We consider the Kerr metric $\tilde h_{ij}$ given by (\ref{eq:kerr-metric}),
on the manifold $M=\Rt\setminus \{0\}$, where the functions $u$ and
$q$ are given by (\ref{eq:51}). The Ricci scalar of $\tilde h_{ij}$ is
denoted by $\tilde R$ and the Ricci scalar of the the conformal metric
$h_{ij}$ (defined by (\ref{brilldef})) is denoted by $R$. 
We emphasize that all the functions involved are smooth on $M$.
 
\begin{lemma}
\label{t:psi}
Assume $|a|\leq m$ and $m\neq 0$. Then, we have that $u>0$ everywhere on $M$.
\end{lemma}
\begin{proof}
  We prove this by explicitly showing that $\psi >1$ if $m\neq 0$,
  where $\psi=1+u$. Using (\ref{eq:51}) and (\ref{eq:X2}) we obtain
\begin{equation}
\label{eq:psiX}
 \psi^4 \geq \frac{\tilde r^2}{r^2}. 
\end{equation}
We use equation (\ref{eq:rtilde}) for $\tilde r$   to obtain
\begin{equation}
  \label{eq:5}
  \frac{\tilde r}{r}=1+\frac{m}{r}+\frac{m^2-a^2}{4r^2}\geq 1+\frac{m}{r}.
\end{equation}
Using (\ref{eq:psiX}) and  (\ref{eq:5}) we get our final inequality
\begin{equation}
  \label{eq:10}
   \psi \geq \sqrt{1+\frac{m}{r}}.
\end{equation}. 

\end{proof}

We also mention  the following upper bound for $\psi$  
that can be obtained from equation (\ref{eq:X1})
\begin{equation}
  \label{eq:8}
  \psi^2 \leq \frac{\tilde r }{r} \left(1+\frac{a^2}{\tilde r^2}
  \right),
\end{equation}
for the extreme case  $|a|=m$ this reduces to 
\begin{equation}
  \label{eq:23}
   \psi^2 \leq \left(1+\frac{m}{r}\right)\left(1+\frac{m^2}{(r+m)^2}\right) .
\end{equation}
These bounds are sharp at infinity, in the sense that the right hand
side of (\ref{eq:8}) and (\ref{eq:23}) goes to $1$ as $r\to
\infty$. 

\begin{lemma}
\label{lemmaRtildePositivo}
Assume $|a|\leq m$ and $a\neq 0$. Then, $\tilde R \geq 0$ on $M$, and
$\tilde R=0$ only at $\rho=0$.
\end{lemma}
\begin{proof}
  The first statement is immediate if one considers the Hamiltonian
  constraint equation in the given slice and the fact that these
  slices are maximal (i.e. the trace of the second fundamental form is
  zero). In effect, since
  $\tilde K=\tilde K_{ij} \tilde h^{ij}=0$  (where $\tilde K_{ij}$ is the
  second fundamental form of the slice) the Hamiltonian constraint is
  given by 
\begin{equation}
  \label{eq:3}
  \tilde R = \tilde K_{ij} \tilde K^{ij}\geq 0.
\end{equation}
To prove the second part of the theorem we compute the right hand side
of (\ref{eq:3}). 
The Kerr second fundamental form $\tilde K_{ij}$ can be expressed in
terms of derivatives of a potential $\omega$  (see, for example, 
\cite{Dain05d}\cite{Dain05c}). In particular, for the square of $K_{ij}$
we have
\begin{equation}
\label{eq:KK}
 \tilde K_{ij} \tilde K^{ij}=\frac{e^{-2q}|\partial
   \omega|^2}{2 \psi^{4}X^2}.  
\end{equation}
The explicit expression for $\omega$ is given in equation
(\ref{eq:omega}).  We calculate 
\begin{equation}
\label{eq:oo}
|\partial \omega|^2 =
\frac{4m^2a^2\rho^6}{r^8\Sigma^4} F,  
\end{equation}
where
\begin{equation}
  \label{eq:14}
   F=\left( 4\Delta a^4\rt^2(\sin(2\theta))^2
+\left(3\rt^4+a^2\rt^2+a^2(\rt^2-a^2)\cos^2\theta\right)^2\right).
\end{equation}
We have
\begin{equation}
  \label{eq:15}
  F\geq 9 \tilde r^8.
\end{equation}
Using (\ref{eq:KK}) and (\ref{eq:oo}) and the explicit expressions
(\ref{eq:51}) we obtain the following lower bound
\begin{equation}
  \label{eq:11}
  \tilde R \geq \frac{18 m^2 a^2 \tilde r^8\sin^2\theta}{(\tilde r^2 +a^2)^7},
\end{equation}
which proves the theorem because $\tilde r \geq m$. 
Note that for the extreme case this bound reduces to
\begin{equation}
  \label{eq:12}
  \tilde R \geq \frac{18 m^{4} (r+m)^8 \sin^2\theta}{\left( (r+m)^2
      +m^2\right)^7}.
\end{equation}
As a side remark we point out that the bounds (\ref{eq:11}) and
(\ref{eq:12}) are sharp in the limit $r\to 0$ and $r\to \infty$.

\end{proof}

\begin{figure}
\centering
 \includegraphics[height=4cm]{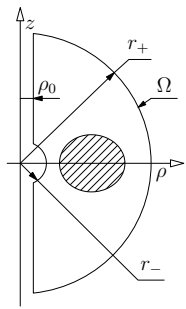} 
 \caption{Non-extreme case: the region $\Omega$ contains the support
   of $R^-$ (shaded region).} \label{fignoext}
\end{figure}
For the next lemma we define the function $R^-$ to be the negative
part of $R$
\begin{equation}
R^- =\min\{0,R\}.
\end{equation}

\begin{lemma}\label{lemmaRmenosCompacto}
Assume $|a|<m$. Then the function $R^-$ has compact support in
$M$. Moreover, the support of $R^-$  does not
intersect the  $\rho=0$ axis (see figure \ref{fignoext})
\end{lemma}

\begin{proof}
  Take a small ball centered at the origin. The leading order term in
  the asymptotic expansion of $R$ for $r\to 0$ (see equation
  (\ref{eq:TnoEa})) is positive. Hence, there exists a constant $r_-$
  such that the function $R$ is positive in small ball of radius $r_-$
  centered at the origin. We also have that $R$ is positive evaluated
  at the axis $\rho=0$ (see equation (\ref{eq:posR})). Then, there
  exist a small cylinder of radius $\rho_0$ such that $R$ is positive
  outside this cylinder.

  In a similar fashion, the leading order term in the asymptotic
  expansions of $R$ for $r\to \infty $ (see equation (\ref{eq:TnoE})) is
  positive. Hence, there exist a constant $r_+$ such that $R$ is
  positive outside a ball of radius $r_+$. Bringing these together, we
  conclude that the support of $R^-$ is contained in the region
  $\Omega$ given by
\begin{equation}\label{E}
 \Omega=\lbrace  r_-<r<r_+ \quad \cap \quad \rho_0<\rho\rbrace 
\end{equation} 
The region $\Omega$ is showed in Fig. \ref{fignoext}, where we also show 
how the actual support of $R^-$ looks like when obtained by explicit 
computation for some values of $a$ and $m$.
\end{proof}

The region $\Omega$ defined above will be used in the next section in
the following way. One point, let say $s_1$, is located at the origin
of $\Omega$. The other point $s_2$ is chosen to be outside $\Omega$,
that is $d > r_+$. Let $\psi_2$ the conformal factor of a Kerr metric
with respect to $s_2$. Then, the upper bounds (\ref{eq:8})--(\ref{eq:23})
imply that  $\psi_2$ is bounded in $\Omega$ (note that $\psi_2$ is not
bounded at $s_2$). For points in the interior of $\Omega$ we have
\begin{equation}
  \label{eq:6}
  r_2\geq d-r\geq d-r_+,
\end{equation}
and then the upper bound  (\ref{eq:8}) in $\Omega$  gives the
following bound which depends only on $d$ and $r_+$
\begin{equation}
  \label{eq:7}
  \psi_2^2 \leq\left(1+\frac{1}{(d-r_+)} +\frac{m^2-a^2}{4(d-r_+)^2}
  \right)\left(1+\frac{a^2}{(d-r_++m)^2 }
  \right).
\end{equation}
For the extreme case this bound reduces to
\begin{equation}
  \label{eq:13}
   \psi_2^2 \leq\left(1+\frac{1}{(d-r_+)}
  \right)\left(1+\frac{m^2}{ (d-r_++m)^2 }
  \right).
\end{equation}
The important point is that in the limit $d\to \infty$ we have
$\psi_2\to 1$ in $\Omega$ (and hence $u_2\to 0$ in $\Omega$).

\begin{figure}
 \centering
 \includegraphics[height=4cm]{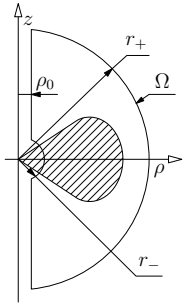}
  \caption{Extreme case: the support of $R^-$ (shaded region) is not
    contained in the region $\Omega$.}  \label{fig-ext}
\end{figure}

The statement of lemma \ref{lemmaRmenosCompacto} is false for the
extreme case $|a|=m$. The reason for this is that $R$ has a different 
behavior at the origin in that case (see equation
(\ref{eq:TE})), which reflects the change in this limit from an
asymptotically flat end to a cylindrical one.  In figure \ref{fig-ext}
we show the support of $R^-$ in the extreme case. We see that the
support of $R^-$ `touches' the origin and hence lies outside of the
region $\Omega$. This is the main technical difficulty  to prove theorem
\ref{t:main} in the extreme limit. To analyze that case we compute the
following formula for $R$.
\begin{lemma}
\label{RRE}
  Let $|a|=m$. Then, the scalar curvature $R$ is given by
  \begin{equation}
    \label{eq:9}
    R= e^{-2q} \left ( -
\frac{3|\partial \omega|^2}{2X^2}+ \frac{8|\partial \psi|^2}{\psi^2} \right).
  \end{equation}
\end{lemma}
\begin{proof}
  Define $\sigma$ by $\psi^4=e^\sigma$. The metrics $\tilde h_{ij}$
  and $h_{ij}$ are related by the conformal transformation $\tilde
  h_{ij}= e^\sigma h_{ij}$, hence the  relation between the corresponding
  scalars curvatures $\tilde R$ and $R$ is given by
  \begin{equation}
    \label{eq:16}
   R= \tilde R e^\sigma +2e^{-2q}\Ldt \sigma + \frac{1}{2}e^{-2q} 
    |\partial \sigma|^2,
  \end{equation}
where $\Ldt$ is the flat Laplacian in 3-dimensions. 
The important step is that the Kerr metric satisfies the stationary and
axially symmetric equations. For the extreme case, these
equations imply (see \cite{Dain05c})
\begin{equation}
  \label{eq:17}
 \Ldt \sigma=- \frac{|\partial \omega|^2}{X^2}. 
\end{equation}
We emphasize that in our coordinates equation (\ref{eq:17}) is valid only for
the extreme case (see the discussion in \cite{Dain05c}). 

Combining (\ref{eq:16}), (\ref{eq:17}) and the expression for $\tilde
R$ which comes from the Hamiltonian constraint equation (\ref{eq:3})
and (\ref{eq:KK}) the result follows.  
\end{proof}

\section{Proof of the main result}
The important feature of the metric (\ref{brillsuma}) is that its
Ricci scalar has a simple decomposition in terms of the Ricci scalars
of each individual Kerr metric, namely 
\begin{equation}
\label{Rtot}
R_{\tilde H}=
\frac{\psi_1^5e^{-2q_2}}{\Psi^5} \left(\tilde R_1+\frac{R_1}{\psi_1^5}u_2\right)+
\frac{\psi_2^5e^{-2q_1}}{\Psi^5} \left(\tilde R_2+\frac{R_2}{\psi_2^5}u_1\right),
\end{equation}
where  $\Psi=1+u_1+u_2$,  $\psi_1=1+u_1$ and  $\psi_2=1+u_2$.  
Note that the two terms in (\ref{Rtot}) are symmetric with respect to
the points labels $1$ and $2$. 

For the non-extreme case the formula (\ref{Rtot}), together with
previous lemmas, gives the desired result. 
\begin{theorem}[Non-extreme case]
\label{TeoremaNoExtremo}
Assume that $|a_1|<m_1$, $|a_2|<m_2$. Then there exists a constant
$d_c$ such that if $d>d_c$, then $R_{\tilde H} \geq 0$. Moreover, if $a_1$
or $a_2$ is different from zero then we have $R_H > 0$ on $S$ for $d>d_c$.
\end{theorem}

\begin{proof}
To analyze the sign of $R_{\tilde H}$ we study each of the two terms in
(\ref{Rtot}) separately. Since they are symmetric in $1$ and $2$ it is
enough to analyze only one of them. 
If $a_1=a_2=0$ we have $R_{\tilde H}=0$ (see equation (\ref{eq:Schw}))
and then the conclusion of the theorem follows. Hence,
in the following we will assume that either $a_1$ or $a_2$ is
different from zero. Without loss of
generality we take  $a_1\neq 0$ and analyze the first term in
(\ref{Rtot}). 

By lemma \ref{t:psi} the factor which multiplies the parenthesis on that
term  is positive
definite. Then, we only need to analyze the sign of
\begin{equation}
\label{eq:K1}
 K_1:=\tilde R_1+\frac{R_1}{\psi_1^5} u_2.
\end{equation}
Take a fixed coordinate system $(\rho,z, \phi)$ centered at the point
$s_1$. In these coordinates we have $s_2$ located at $z=-d$. Note that, 
in these coordinates, the only function in (\ref{eq:K1}) which depends on $d$
is $u_2$.
 
Using lemma \ref{t:psi} and \ref{lemmaRtildePositivo} we known that
the only function which is not positive in the definition of $K_1$ is
$R_1$. Let $\Omega$ be the region defined in lemma
\ref{lemmaRmenosCompacto} centered at $s_1$, with constants
$(\rho_0,r_-, r_+)$. Take $d$ such that $d > r_+$. That is, the point
$s_2$ is not in $\Omega$. Since $R_1$ is positive outside $\Omega$, 
it follows that $K_1$ is positive outside $\Omega$. Note that the
constants $(\rho_0,r_-, r_+)$  in  (\ref{E}) do not depend on $d$. 

We now turn our attention to the interior of $\Omega$.  By lemma
\ref{lemmaRtildePositivo} we know that $\tilde R_1$ is strictly
positive in $\Omega$.  We use the upper bound (\ref{eq:7}) for $u_2$
in the interior of $\Omega$. We have that $u_2\to 0$ in $\Omega$ as
$d\to \infty$.  Since $\tilde R_1$ and $R_1$ do not depend on $d$ 
% and
% $\tilde R$ is strictly positive in $\Omega$, 
it follows that there exists a
constant $d_1$ such that $K_1$ is positive in the interior of $\Omega$
for all $d>d_1$.

For $K_2$ we repeat this analysis to obtain a constant $d_2$ such that
$K_2 > 0$ is positive for all $d> d_2$. We chose $d_c$ to be the maximum
between $d_1$ and $d_2$, from which the conclusion of the theorem follows. 
\end{proof}

The proof in the extreme case is more delicate because lemma
\ref{lemmaRtildePositivo} is no longer valid.  However, a more detailed 
analysis of the asymptotic expansion at the origin of the function $K_1$, 
using lemma \ref{RRE}, allows the desired extension.

\begin{theorem}[Extreme case]
\label{TeoremaExtremo}
Assume that either $|a_1|=m_1$, $|a_2| <m_2$ or $|a_2|=m_2$, $|a_1|
<m_1$ or $|a_1|=m_1$, $|a_2| = m_2$ . 
Then there exists a constant
$d_c$ such that if $ d>d_c$ then $ R_H\geq 0$ everywhere on
$S$. Moreover, if $a_1$ or $a_2$ are different from zero, then 
$ R_H > 0$  on $S$ for $ d>d_c$.
\end{theorem}

\begin{proof}
  If $a_1=a_2=0$ then the result is trivial since $R_{\tilde H}=0$ in
  that case. We assume that $|a_1|=m_1$ and $a_1\neq 0$. We make no
  restriction on the parameter $a_2$, that is $0\leq |a_2| \leq m_2$.
  By the symmetry in $1$ and $2$ of (\ref{Rtot}) this assumption will
  cover all cases.
As in the non-extreme case, we use a coordinate system centered at
$s_1$.  

We analyze $K_1$ directly. We have that  $R_1$ is positive outside some
ball of large radius  $r_+$ (because
the leading term in the asymptotic expansion (\ref{eq:TE}) is
positive). As in the previous theorem we chose $d$ such that
$d>r_+$. From equation (\ref{eq:K1}) it follows that $K_1$ is positive
outside the ball $B$ of radius $r_+$ centered at $s_1$. To analyze the
behavior of $K_1$ in  the interior of   $B$ 
we use the explicit expression for $R_1$ given
by lemma \ref{RRE} and the explicit form of $\tilde R_1$ given by
(\ref{eq:3}) and (\ref{eq:KK}) to obtain
\begin{equation}
  \label{eq:26}
  K_1=\frac{e^{-2q_1}}{\psi_1^4}
 \left(  \frac{|\partial\omega_1|^2}{2X_1^2}\left(1-
     \frac{3u_2}{\psi_1} \right) + 
\frac{8|\partial\psi_1|^2}{\psi^2_1}\frac{u_2}{\psi_1} 
\right) 
\end{equation}
This expression is clearly positive if
\begin{equation}
  \label{eq:27}
 \frac{u_2}{\psi_1} <\frac{1}{3}.
\end{equation}
By (\ref{eq:13}) we have that $u_2$ is bounded in $B$ and it goes to
zero as $d\to \infty$. Then, condition (\ref{eq:27}) can always be
achieved for sufficiently large $d$. Note that using the upper bound
(\ref{eq:13}) the constant $d_c$ can be explicitly calculated. We have
proved that $K_1\geq 0$. To prove that it is strictly positive, we use
that $|\partial \omega|^2$ only vanishes at the axis (see equation
(\ref{eq:oo}), (\ref{eq:14}) and (\ref{eq:15})) and at the axis we can
explicitly compute the term $|\partial \psi_1|^2$ to see that it is
strictly positive.

\end{proof}

\section{Final comments}

% Sum more than two
We have constructed our metric as a sum of two Kerr black hole
metrics. It is straightforward to generalize this result to include more
than two black holes, that is to superpose more $u_k$ and $q_k$ in the 
definition of the conformal metric (\ref{brillsuma}).

% The general case
We have assumed that the conformal metric is axially symmetric. Due to
the particular simpler expression of the scalar curvature of this
metric (see equation (\ref{Rtot})), this assumption represents a mayor
simplification with respect to the case where the spins point in
arbitrary directions.  We expect that a  result similar to theorem
\ref{t:main} remains true in the general case. However, proving such a
result requires estimations for many new terms in the scalar curvature
and it is not clear how to do this in an efficient way.

% Superposition of Brill waves
Even in the axially symmetric case, it is interesting to see what kind
of properties of the individual metrics are necessary to make the
superposition. We have provided sufficient conditions which are
satisfied by the Kerr metric. We do not known if these properties are
also necessary. Our first attempt was to consider the superposition of
two arbitrary Brill metrics (i.e metrics of the form (\ref{brilldef}))
with positive Yamabe invariant. But we were unable to prove that an
analog of theorem \ref{t:main} holds under these weak
assumptions. Inspired by the Kerr data, our next attempt was to impose
that the negative part of the Ricci scalar $R^-$ has compact
support. This seems to be a natural assumption, because the support of
$R^-$ seems to play the role of the `body' and the separation can be
the taken as the distance between these sets. Furthermore, for Brill
metrics the support of $R^-$ is always non trivial because the
integral of the Ricci scalar over $\Rt$ is zero (see \cite{Brill59}).
However, for the proof we have also used another particular property
of the Kerr metric, namely $u>0$. It is not clear to us whether this
condition is actually necessary.

\section*{Acknowledgments}
It is a pleasure to thank Robert Geroch for discussions.  S. Dain is
supported by CONICET (Argentina). G. Avila was partially supported by a
`Conciencias' fellowship, from the Agencia C\'ordoba Ciencia
(Argentina).  This work was supported in part by grant PIP 6354/05 of
CONICET (Argentina), grant 05/B270 of Secyt-UNC (Argentina) and the
Partner Group grant of the Max Planck Institute for Gravitational
Physics, Albert-Einstein-Institute (Germany).

\appendix
\section{Asymptotic expansions for Kerr initial data}
\label{sec:kerr-data}
In this section we study the behavior of Kerr initial data (with
parameters $m$ and $a$) in the limits $r\to 0$, $r \to \infty$ and
$\rho \to 0$.

The explicit expressions for the functions $q$ and $u$, which
characterize the Kerr intrinsic metric (for Boyer-Lindquist slices)
$\tilde h_{ij}$, in terms of the coordinates $(\rho,z,\phi)$ are given
by (see, for example,\cite{Dain05c})
\begin{equation}
  \label{eq:51}
  \psi^4= \frac{X}{\rho^2}, \quad
  e^{2q}=\frac{\Sigma\sin^2\theta}{X},\quad u=\psi -1,
\end{equation}
where
\begin{equation}
 \Delta 	=\tilde r^2+a^2-2m\tilde r, \quad 
 \Sigma 	=\tilde r^2+a^2 \cos^2 \theta,
\end{equation}
 and  
\begin{align}
X &=\left[\frac{(\tilde r^2+a^2)^2 -\Delta a^2 \sin^2\theta}{\Sigma}
\right]\sin^2\theta, \label{eq:X1}\\
& = \left (\tilde r^2+a^2+\frac{2m\tilde r a^2 \sin^2
    \theta}{\Sigma}\right) \sin^2 \theta. \label{eq:X2}
\end{align}
Where have defined $\rho=r\sin\theta$,  $z=r\cos\theta$ and
\begin{equation}
\label{eq:rtilde}
\tilde r = r +m+\frac{m^2-a^2}{4 r}.
\end{equation}

The potential $\omega$ which characterizes the Kerr second fundamental
form $\tilde K_{ij}$ is given by (see, for example, \cite{Dain05c})
\begin{equation}
\label{eq:omega}
\omega = 2ma(\cos^3\theta-3\cos\theta)-
\frac{2ma^3\cos\theta\sin^4\theta}{\Sigma}.  
\end{equation}

In the following $\tilde R$ denotes the Ricci scalar of $\tilde
h_{ij}$ and $R$ the Ricci scalar of the conformal metric $h_{ij}$ defined
by (\ref{brilldef}).

In the limit $a=0$ we have Schwarzschild data
\begin{equation}
  \label{eq:Schw}
  \psi= 1+\frac{m}{2r}, \quad q=0,\quad R=0, \quad \tilde R =0. 
\end{equation}

In the following, we assume $a\neq 0$. 
We analyze first the non-extreme case $|a|<m$. In this case, $u$ has the
following asymptotic behavior
\begin{align}
 \label{AppPsiNoExt}
   u & =\frac{m}{2r}+\orden{r^{-2}} \text{ as }r\to \infty,  \\
 \label{eq:53}
 u & = \frac{\sqrt{m^2-a^2}}{2r}+\orden{1}  \text{ as }r\to 0.
\end{align} 

The scalar curvature $R$ satisfies
\begin{align}
 \label{eq:TnoE} 
R & =\frac{2a^2}{r^4}+\orden{r^{-5}}  \text{ as }r\to \infty,  \\
R & =2\left(\frac{4a}{m^2-a^2}\right)^2+\orden{r^2}  \text{ as }r\to
0. \label{eq:TnoEa}
\end{align}

For the extreme case $a=m$ we have that at infinity $u$ has the
same behavior as in the non-extreme case, but at the origin this
changes to 
\begin{align}
 \label{eq:psiE}
 u & = \frac{\sqrt{2m}}{\sqrt{r}(1+\cos^2\theta)^{1/4}}+\orden{r^{-1}}
 \text{ as }r\to 0
\end{align}

The behavior of the scalar curvature $R$ and $\tilde R$ in this case
is given by
\begin{align}
  R & =\frac{2m^2}{r^4}+\orden{r^{-5}}    \text{ as }r\to \infty\\
\label{eq:TE}
  R &
  =16\left[\frac{3\cos^2\theta-1}{(1+\cos^2\theta)^4}\right]\frac{1}{r^2}
  + \orden{r^{-1}}  \text{ as }r\to 0.
\end{align}
and
\begin{equation}
  \label{eq:RtildeE}
\tilde R = \frac{2\sin^2(\theta)}{m^2(1+\cos^2(\theta))} +\orden{r} \text{ as }r\to 0.
\end{equation}

Finally, we analyze the behavior of $R$ near the axis $\rho=0$. For a
Brill metric like (\ref{brilldef}), the Ricci scalar is given by
\cite{Brill59}  
\begin{equation}
\label{eq:Rb}
 R=-2e^{-2q}\Ld q,
\end{equation}
where $\bar \Delta$ is the flat Laplacian in two dimensions, which in
coordinates $r,\theta$ is given by
\begin{equation}
 \Ld = \frac{1}{r}\partial_r
 (r \partial_r)+\frac{1}{r^2} \partial^2_\theta. 
\end{equation}
The function $q$ defined by (\ref{eq:51}) vanishes at the axis
\begin{equation}
  \label{eq:1}
  q(\rho=0,z)=0. 
\end{equation}
This can be of course explicitly seen from (\ref{eq:51}).
It is also a consequence of the regularity of the metric
(\ref{brilldef}) at the axis (see \cite{Brill59}). As a consequence of
(\ref{eq:1}), we have that the radial derivatives of $q$ evaluated at
$\rho=0$ (which are equivalent to $z$ derivatives evaluated at $\rho
=0$, and hence tangential to the axis where $q$ is constant) vanish.
Then, in order to calculate $R$ using the formula (\ref{eq:Rb}) we
need to compute only the derivatives in $\theta$. 
To calculate them, it is convenient to make the change of variable
$\e=2\theta$.  We have  $\partial^2_\theta =4\partial^2_\epsilon $ 
and  $2\cos^2\theta=1+\cos \e$. Then we obtain
\begin{equation}
 A:=e^{2q}=\gamma \frac{(\alpha+\cos\e)^2}{(\beta+\cos\e)},
\end{equation}
where the following are $\gamma$, $\alpha$ and $\beta$ are functions
of $\rt$ given by
\begin{align}
 \alpha&:=2\frac{\rt^2}{a^2}+1, \\
 \beta&:=2\frac{(\rt^2+a^2)^2}{a^2\Delta}-1, \\
 \gamma&:=\frac{a^2}{2\Delta}.
\end{align}
Calculating the required derivatives we get
\begin{equation}
 \frac{1}{A}\frac{\partial A}{\partial \e}
=-\sin\e\left[\frac{2}{\alpha+\cos\e}-\frac{1}{\alpha+\cos\e} \right].
\end{equation}
and
\begin{multline}
\label{eq:sdA}
 \frac{1}{A}\frac{\partial^2 A}{\partial \e^2}
=-\cos\e\left[\frac{2}{\alpha+\cos\e}-\frac{1}{\alpha+\cos\e} \right]+
\sin^2\e\left[\frac{2}{\alpha+\cos\e}-\frac{1}{\alpha+\cos\e} \right]^2- \\
-\sin^2\e\left[\frac{2}{(\alpha+\cos\e)^2}-\frac{1}{(\alpha+\cos\e)^2} \right]^2
\end{multline}
Using (\ref{eq:sdA}) we finally obtain 
\begin{equation}
\label{eq:qtt}
  \frac{\partial^2 q}{\partial
    \e^2}=-\left[\frac{1+\alpha\cos\e}{(\alpha+\cos\e)^2}\right]+ 
  \left[\frac{1+\beta\cos\e}{(\beta+\cos\e)^2}\right].
\end{equation}
 Using (\ref{eq:Rb}) and (\ref{eq:qtt}) evaluating at the axis  $\e=0$
 we get
\begin{equation}
  R(\rho=0)= \frac{-4}{r^2}\left[\frac{1}{1+\beta}-\frac{2}{1+\alpha}\right].
\end{equation}
To prove that this expression is positive, 
consider the following inequality
\begin{equation}
 \beta=\frac{2(\rt^2+a^2)^2}{a^2(\rt^2+a^2-2m\rt)}-1 >
 2\frac{\rt^2}{a^2}+1=\alpha 
\end{equation}
obtained by replacing the denominator $(\rt^2+a^2-2m\rt)$ by
$(\rt^2+a^2)$. Note that we assume $m>0$ in order to get the strict
inequality. This inequality implies 
\begin{equation}
\frac{1}{1+\beta}-\frac{2}{1+\alpha}<0
\end{equation}
and therefore we get our final result
\begin{equation}
\label{eq:posR}
 R(\rho=0)>0,\quad \text{ on } M.
\end{equation}
We emphasize that (\ref{eq:posR}) is valid for both the non-extreme
and extreme cases.

%\bibliographystyle{/home/dain/biblio/habbrv}
%\bibliography{/home/dain/biblio/biblio}

\end{document}